\documentclass[aps,prl,twocolumn,showpacs,preprintnumbers,amsmath,amssymb,superscriptaddress]{revtex4}%

\usepackage{graphicx}%
\usepackage{dcolumn}
\usepackage{amsmath}
\usepackage{color}
\usepackage{bm}

\begin{document}

\preprint{PREPRINT (\today)}

\title{Oxygen-isotope effect on the superconducting gap in the
cuprate superconductor Y$_{1-x}$Pr$_x$Ba$_2$Cu$_3$O$_{7-\delta}$}

\author{R.~Khasanov}
 \affiliation{Physik-Institut der
Universit\"{a}t Z\"{u}rich, Winterthurerstrasse 190, CH-8057
Z\"urich, Switzerland}
\author{S.~Str\"assle}
\affiliation{Physik-Institut der Universit\"{a}t Z\"{u}rich,
Winterthurerstrasse 190, CH-8057 Z\"urich, Switzerland}
\author{K.~Conder}
 \affiliation{Laboratory for Developments and Methods, Paul Scherrer Institute,
CH-5232 Villigen PSI, Switzerland}
\author{E.~Pomjakushina}
 \affiliation{Laboratory for Developments and Methods, Paul Scherrer Institute,
CH-5232 Villigen PSI, Switzerland}
 \affiliation{Laboratory for Neutron Scattering, Paul Scherrer Institute \& ETH
 Zurich,CH-5232 Villigen PSI, Switzerland}
\author{A.~Bussmann-Holder}
 \affiliation{Max-Planck-Institut f\"ur Festk\"orperforschung,
Heisenbergstrasse 1, D-70569 Stuttgart, Germany}
\author{H.~Keller}
 \affiliation{Physik-Institut der Universit\"{a}t Z\"{u}rich,
Winterthurerstrasse 190, CH-8057 Z\"urich, Switzerland}

\begin{abstract}
The oxygen-isotope ($^{16}$O/$^{18}$O) effect (OIE) on the
zero-temperature superconducting energy gap $\Delta_0$ was studied
for a series of Y$_{1-x}$Pr$_x$Ba$_2$Cu$_3$O$_{7-\delta}$ samples
($0.0\leq x\leq0.45$). The OIE on $\Delta_0$ was found to scale with
the one on the superconducting transition temperature. These
experimental results are in quantitative agreement with predictions
from a polaronic model for cuprate high-temperature superconductors
and rule out approaches based on purely electronic mechanisms.
\end{abstract}

\maketitle


Within BCS theory the isotope-effect exponents of the
superconducting transition temperature ($T_c$) and of the
zero-temperature superconducting energy gap ($\Delta_0$) are
constant and have the same values for both quantities. Strikingly
different observations were reported for cuprate high-temperature
superconductors (HTS's), where the isotope-effect on $T_c$ is
strongly doping dependent. The isotope-effect exponent is
vanishingly small at optimum doping but increases nonlinearly when
approaching the underdoped regime
\cite{Franck91Franck94,Zech94,Zhao97,Khasanov04a,Keller05} where it
even exceeds the BCS value of 0.5. Further additional distinctions
arise in HTS's caused by the fact that the superconducting order
parameter is complex with a leading $d-$wave component in
coexistence with a smaller $s-$wave one
\cite{Khasanov07_La214,Kohen03,Muller02,Khasanov07_Y124} as
suggested early on \cite{Muller95,Muller97}. This intricacy does not
allow to draw specific conclusions about any isotope effect on the
zero-temperature superconducting gap $\Delta_0$, especially when
considering that a $d-$wave gap could be a consequence of electronic
effects. If high temperature superconductivity were caused by strong
electronic correlations only, no isotope effect on the
zero-temperature gap is expected. On the other hand, the coexisting
$s-$wave gap could originate from electron-lattice interactions and
carry an isotope effect. Thus, a methodic investigation of an
isotope effect on the complex superconducting gap together with its
relation to the one on $T_c$ admits to conclude about the nature of
the pairing mechanism.

In this paper we report the studies of the oxygen-isotope
($^{16}$O/$^{18}$O) effect (OIE) on the zero-temperature
superconducting energy gap $\Delta_0$ in the cuprate superconductor
Y$_{1-x}$Pr$_x$Ba$_2$Cu$_3$O$_{7-\delta}$. A linear relation between
$\Delta_0$ and $T_c$ is found as predicted theoretically. The
isotope effect on $\Delta_0$ scales linearly with the one on $T_c$
and reverses sign around optimum doping, as anticipated  from model
calculations. Different doping levels of the isotope exchanged
samples were ruled out by performing careful back exchange
experiments.

Polycrystalline samples of Y$_{1-x}$Pr$_x$Ba$_2$Cu$_3$O$_{7-\delta}$
($x=0.0$, 0.2, 0.3, 0.45) were prepared by standard solid state
reaction \cite{Conder01}. In order to obtain fine grains needed for
the determination of the magnetic field penetration depth
($\lambda$) by low field magnetization measurements, the ceramic
samples were first grounded for approximately 20 minutes in air and
then passed through 10~$\mu$m sieves. The oxygen-isotope exchange
was performed by heating the samples in $^{18}$O$_2$ gas.
In order to ensure the same thermal history of the substituted
($^{18}$O) and not substituted ($^{16}$O) samples, both annealings
(in $^{16}$O$_2$ and $^{18}$O$_2$ gas) were always made
simultaneously.
The $^{18}$O content was determined from the weight change and found
to correspond to a 90(2)\% exchange. The samples containing Pr were
$c-$axis oriented in a field of 9~T whereas samples with no Pr
remained in non-oriented powder form.

AC and DC magnetization experiments were carried through in the
temperature range 2--100~K by using Quantum Design Magnetometers
(MPMS and PPMS). The samples with no Pr were studied in DC
experiments with a DC field amplitude of 0.5~mT. The oriented
samples containing Pr were investigated with the AC field (field
amplitude 0.3~mT and field frequency 333~Hz) applied parallel to the
$c-$axis. The separation of the grains and the absence of weak links
was tested by confirming the linear relation between the
magnetization and the field at $T=10$~K. The DC field variation
ranged from 0.5~mT to 1.5~mT, and the AC fields from 0.1~mT to 1~mT
with frequencies between 49 and 599~Hz. The magnetization data were
corrected by subtracting the paramagnetic background.

\begin{figure}[htb]
\includegraphics[width=0.9\linewidth]{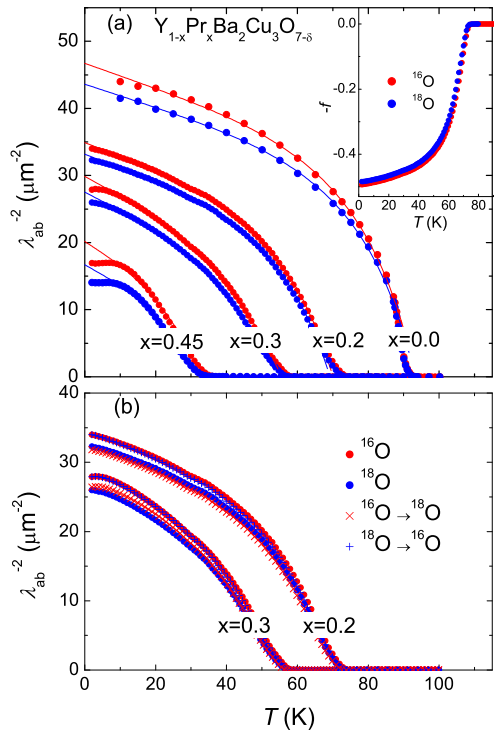}
 \vspace{-0.4cm}
\caption{(Color online) (a) Temperature dependences of the in-plane
magnetic field penetration depth $\lambda_{ab}$ for
$^{16}$O/$^{18}$O substituted
Y$_{1-x}$Pr$_x$Ba$_2$Cu$_3$O$_{7-\delta}$ ($x=0.0$, 0.2, 0.3, 0.45)
samples. The solid lines are results from a numerical analysis as
described in the text.  The inset shows the temperature dependences
of the Meissner fraction $f$ for $^{16}$O/$^{18}$O substituted
Y$_{0.8}$Pr$_{0.2}$Ba$_2$CuO$_{7-\delta}$. (b) $\lambda_{ab}(T)$ for
back exchanged ($^{16}$O$\rightarrow$$^{18}$O,
$^{18}$O$\rightarrow$$^{16}$O)
Y$_{1-x}$Pr$_x$Ba$_2$Cu$_3$O$_{7-\delta}$ ($x=0.2$, 0.3), showing
that the doping level for the $^{16}$O/$^{18}$O samples is
unchanged. }
 \label{fig:lambda}
\end{figure}

The Meissner fractions ($f$) were obtained from the measured
magnetization, sample masses, and their X-ray density inferring a
spherical grain shape. In the inset of Fig.~\ref{fig:lambda}~(a)
$f(T)$ is shown for the $c-$axis aligned $^{16}$O and $^{18}$O
samples with $x=0.2$. Since the Meissner fraction is substantially
smaller than 1 it is ensured that the average size of the grains is
comparable to $\lambda$. This strong reduction in $f$ is a
consequence of the surface-field penetration in each individual
grain. In addition, in the whole temperature range $f$ is
systematically larger in the $^{16}$O samples than in the $^{18}$O
samples confirming that the penetration depth is reduced in the
former samples as compared to the latter ones, as observed
previously \cite{Zhao97,Khasanov04a,Hofer00,Khasanov03,Khasanov04}.
The temperature dependence of $\lambda$ was derived within the
Shoenberg model \cite{Shoenberg40}.
The following issues are important for the discussion of the
experimental results: (i) When a small magnetic field is applied
along the $c-$axis, the screening currents flow in the $ab-$plane,
decaying on the distance $\lambda_{ab}$  from the grain surface.
Thus experiments on the $c-$axis oriented samples provide direct
information on $\lambda_{ab}$. (ii) From experiments on non-oriented
powders an effective averaged penetration depth ($\lambda_{eff}$)
can be extracted. However, in highly anisotropic extreme type-II
superconductors (as HTS's) this can -- in turn -- be related to the
in-plane penetration depth through the relation  $\lambda_{eff}
=1.23 \lambda_{ab}$ \cite{Fesenko91}. (iii) The absolute value of
$\lambda_{ab}$ depends on the value of the Meissner fraction $f$ and
the average grain size radius \cite{Shoenberg40}, whereas the
temperature dependence is independent of the grain size and entirely
given by $f(T)$ \cite{Panagopoulos96}. (iv) In order to account for
the undetermined average grain size, the values of
$\lambda_{ab}^{-2}(2$~K) for the $^{16}$O samples were normalized to
those obtained from muon-spin rotation ($\mu$SR) experiments
\cite{Seaman90}.

\begin{table*}[htb]
\caption[~]{\label{Table:results} Values of $T_c$ and $\Delta_0$ for
the $^{16}$O/$^{18}$O substituted
Y$_{1-x}$Pr$_x$Ba$_2$Cu$_3$O$_{7-\delta}$ samples investigated in
this work. The relative isotope shifts of these quantities are
displayed in the last two columns of the table. }
\begin{center}
\begin{tabular}{lcc|cc|ccccccc} \hline\hline
&\multicolumn{2}{c}{$^{16}$O}&\multicolumn{2}{c}{$^{18}$O}&\multicolumn{2}{c}{}\\
Sample&$T_c$&$\Delta_0$&$T_c$&
       $\Delta_0$&${\delta T_c}/{T_c}$&
       ${\delta \Delta_0}/{\Delta_0}$\\
&[K]&[meV]&[K]&[meV]&[\%]&[\%]\\
 \hline
 YBa$_2$Cu$_3$O$_{7-\delta}$                 &93.23(7)&29.75(22)&
                                             93.01(6)&30.06(24)&-0.22(11)&1.1(1.2)\\
Y$_{0.8}$Pr$_{0.2}$Ba$_2$Cu$_3$O$_{7-\delta}$&70.02(6)&19.61(14)&
                                             69.22(8)&19.33(13)&-1.25(16)&-1.6(1.1)\\
Y$_{0.7}$Pr$_{0.3}$Ba$_2$Cu$_3$O$_{7-\delta}$&55.50(8)&12.28(9)&
                                             54.40(8)&11.98(11)&-2.16(23)&-2.7(1.3)\\
Y$_{0.55}$Pr$_{0.45}$Ba$_2$Cu$_3$O$_{7-\delta}$&33.01(8)&6.87(5)&
                                             31.20(7)&6.53(5)&-6.06(37)&-5.5(1.2)\\
 \hline
Y$_{0.8}$Pr$_{0.2}$Ba$_2$Cu$_3$O$_{7-\delta}$&69.80(6)\footnotemark[1]&19.54(12)\footnotemark[1]&
                                              69.02(7)\footnotemark[2]& 19.19(13)\footnotemark[2]&-1.25(14)&-1.99(1.0)\\
Y$_{0.7}$Pr$_{0.3}$Ba$_2$Cu$_3$O$_{7-\delta}$&55.41(8)\footnotemark[1]&12.21(8)\footnotemark[1]&
                                              54.30(7)\footnotemark[2]&11.94(8)\footnotemark[2]&-2.21(22)&-2.04(1.0)\\

  \hline \hline
\end{tabular}
 \footnotetext[1]{Results for back-exchanged  $^{18}$O$\rightarrow$$^{16}$O
 samples}
 \footnotetext[2]{Results for back-exchanged  $^{16}$O$\rightarrow$$^{18}$O
 samples}

 \end{center}
\end{table*}

The experimental results for the in-plane magnetic field penetration
depth  $\lambda_{ab}$ for $^{16}$O/$^{18}$O substituted
Y$_{1-x}$Pr$_x$Ba$_2$Cu$_3$O$_{7-\delta}$ with $x=0.0$, 0.2, 0.3,
0.45 are shown in Fig.~\ref{fig:lambda}~(a). In order to guarantee
that for the $^{16}$O and $^{18}$O substituted samples the doping
level remains the same, back-exchange experiments were performed for
two representative compositions [see Fig.~\ref{fig:lambda}~(b)]. It
is important to note here, that these back-exchange experiments are
absolutely essential since they guarantee that any compositional
deviations or preparation errors are excluded. Only in this way {\it
real} isotope effects can be observed in contrast to {\it marginal}
ones caused by different doping levels \cite{Bishop07}.

The analysis of the data presented in Fig.~\ref{fig:lambda}~(a) was
made within the BCS scheme, extended to account for a $d-$wave gap
and using tabulated values for the temperature dependence of the
normalized gap  (see Ref.~\onlinecite{Khasanov07_La214}). For
$T>10$K all experimental data are well described by this approach
with a nearly linear temperature dependence up to $T\approx0.5T_c$.
Below 10~K the experimental data saturate in contrast to the
calculated curves which continue to increase. This deviance between
experiment and theory can be a consequence of impurity scattering
\cite{Hirschfeld93}, chemical and structural defects
\cite{Panagopoulos96a}, or may be due to the above mentioned
simplifying assumption that the gap is of $d-$wave symmetry only.
From the data neither of these sources can be identified
unambiguously. The values of $T_c$ and $\Delta_0$ obtained from the
analysis of $\lambda^{-2}_{ab}(T)$ data are summarized in
Table~\ref{Table:results}.
The relative isotope shifts of $T_c$ and $\Delta_0$ were determined
from their relative percentage change with isotope substitution. The
values of $\delta T_c/T_c$, and $\delta\Delta_0/\Delta_0$ ($\delta
X/X=[^{18}X-$$^{16}X]/$$^{16}X$, $X=T_c$ or $\Delta_0$), corrected
for the incomplete isotope exchange in the $^{18}$O samples, are
summarized in Table~\ref{Table:results}. The results for the OIE on
$T_c$ are in accord with already published  data
\cite{Franck91Franck94,Zech94,
Khasanov04a,Hofer00,Khasanov03,Khasanov04}.

\begin{figure}[htb]
\includegraphics[width=0.9\linewidth]{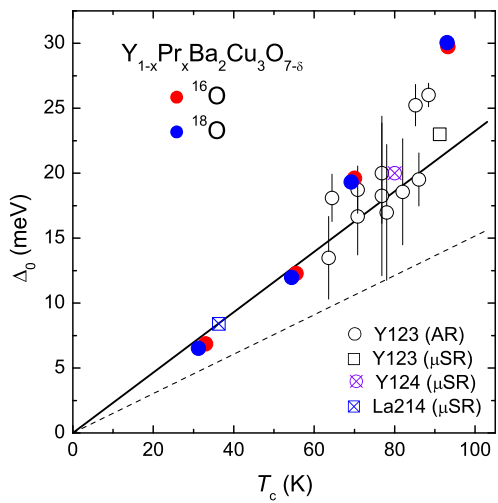}
 \vspace{-0.8cm}
\caption{(Color online) The zero-temperature superconducting gap
$\Delta_0$ {\it vs}. the superconducting transition temperature
$T_c$ for $^{16}$O/$^{18}$O substituted
Y$_{1-x}$Pr$_x$Ba$_2$Cu$_3$O$_{7-\delta}$ samples studied in the
present work and for various hole-doped HTS's studied by means of
Andreev reflection (AR) \cite{Kohen03} and  muon-spin rotation
($\mu$SR) \cite{Khasanov07_La214,Khasanov07_Y124}. The solid line
corresponding to $2\Delta_0/k_BT_c=5.34$ was predicted theoretically
by using a two-component model with polaronic coupling
\cite{Bussmann-Holder07}. The dashed line corresponds to the BCS
value $2\Delta_0/k_BT_c=3.52$  }
 \label{fig:gaps}
\end{figure}

In Fig.~\ref{fig:gaps} the values of the zero-temperature gaps
$\Delta_0$ are shown as a function of their corresponding values of
$T_c$. In order to demonstrate their consistency with earlier
reported  gap values, data for varios HTS's obtained by different
methods \cite{Kohen03,Khasanov07_La214,Khasanov07_Y124} are included
together with theoretical results discussed below
\cite{Bussmann-Holder07, Bussmann-Holder05,Bussmann-Holder05a}. It
should be emphasized that the data presented in Fig.~\ref{fig:gaps}
follow the general trend, namely, $\Delta_0$ scales rather linearly
with $T_c$. The linear relation between $\Delta_0$ and $T_c$ is
absolutely nontrivial, since competing and coexisting energy scales
dominate the phase diagram of HTS. Especially, the pseudogap which
is in many experiments undistinguishable from the superconducting
gap, has given rise to statements controverse to our findings, {\it
i.e.}, that $\Delta_0$ increases with decreasing $T_c$ (see {\it
e.g.} Ref.~\onlinecite{Fischer07} and references therein). The
linear relation between $\Delta_0$ and $T_c$ was predicted by BCS
theory with a value of $2\Delta_0/k_BT_c=3.52$, smaller than
observed here (see Fig.~\ref{fig:gaps}). This linear relation
obviously requires that an isotope effect on $T_c$ results in the
same isotope effect on the superconducting energy gap. However, as
was outlined above, the complex symmetry of $\Delta_0$ and the
unidentified pairing mechanism do not admit any conclusions on the
doping dependence of the isotope effect on the gap.

\begin{figure}[htb]
\includegraphics[width=0.9\linewidth]{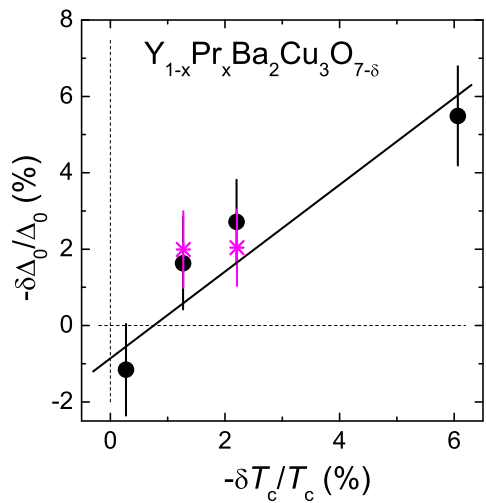}
 \vspace{-0.8cm}
\caption{(Color online) Comparison of the oxygen-isotope shift on
the superconducting gap  $\Delta_0$ with the one on the transition
temperature $T_c$ for Y$_{1-x}$Pr$_{x}$Ba$_2$Cu$_3$O$_{7-\delta}$
($x=0.0$, 0.2, 0.3, 0.45). Circles refer to the present experiments,
the solid line was obtained from model calculations as described in
Refs.~\onlinecite{Bussmann-Holder07,
Bussmann-Holder05,Bussmann-Holder05a}. The stars refer to the
back-exchange data.}
 \label{fig:rel-shifts}
\end{figure}

Even though our values of $\Delta_0$  were derived indirectly and
display some scattering (see Fig.~\ref{fig:gaps}), the OIE on the
gap is {\it independent} of this methodology, since eventual errors
in the absolute values of $\Delta_0$  are systematic due to the same
analysis of the data sets of $^{16}$O and $^{18}$O samples. The OIE
on $\Delta_0$ is compared to the one on $T_c$ in
Fig.~\ref{fig:rel-shifts}, together with theoretically derived
results \cite{Bussmann-Holder07}. Interestingly, the same linear
relation between both is observed in consistency with a model, where
polaronic renormalizations of the single particle energies were
introduced \cite{Bussmann-Holder07,
Bussmann-Holder05,Bussmann-Holder05a}. Of fundamental importance is
the observation of a sign reversal of the isotope effect around
optimum doping as predicted in Refs.~\onlinecite{Bussmann-Holder07,
Bussmann-Holder05,Bussmann-Holder05a}. This novel discovery provides
substantial evidence that polaronic effects control the physics of
HTS.

The theoretical model considers two components where the doped holes
lead to the formation of metallic regions in the otherwise
insulating antiferromagnetic matrix (for details see
Refs.~\onlinecite{Bussmann-Holder07,
Bussmann-Holder05,Bussmann-Holder05a}). These holes couple strongly
to the ionic displacements to form polarons with variable spatial
extent. Since these metallic polaronic clusters carry huge strain
fields, a self organization into patterns (stripes) takes place
which lowers the strain energy and induces interactions between the
matrix and the "polaron stripes" \cite{Bussmann-Holder07}. The
locally strong electron-lattice interaction within the metallic
regions causes an $s-$wave order parameter, in contrast to the
embedding matrix with a $d-$wave order parameter. Interband
interactions between both subsystems guarantee a single transition
temperature together with coupled gaps which were calculated
self-consistently. The important effect of the polaron formation is
an exponential renormalization of the band width which carries an
isotope effect, and an isotope independent level shift.
Correspondingly, all hopping integrals are renormalized, but
contribute in a very different way to the isotope effect
\cite{Bussmann-Holder05a}. By using different values for the
polaronic coupling the average gap
$\Delta_0=\sqrt{\Delta_s^2+\Delta_d^2}$ is attained  as a function
of the corresponding $T_c$ yielding $2\Delta_0/k_BT_c=5.34$. The
results are included in Fig.~\ref{fig:gaps} as straight line. Also
the $\mu$SR \cite{Khasanov07_La214,Khasanov07_Y124} and Andreev
reflection data \cite{Kohen03} diplayed in Fig.~\ref{fig:gaps} refer
to an average gap. The calculated oxygen-isotope shift of the
average gap $\Delta_0$ {\it vs.} the one on $T_c$ is compared with
the present data of Y$_{1-x}$Pr$_x$Ba$_2$Cu$_3$O$_{7-\delta}$ in
Fig.~\ref{fig:rel-shifts}. A good agreement between both, experiment
and theory is observed.  Here it is worth mentioning that the
isotope effect on the individual gaps, {\it i.e}., $\Delta_s$,
$\Delta_d$, is of the same order of magnitude for both gaps, but
always slightly enhanced for the $d-$wave gap as compared to the
$s-$wave one \cite{Bussmann-Holder07}. It is important to emphasize,
that conventional electron-phonon coupling does not lead to this
doping dependent isotope effect but would always yield -- within
weak coupling -- an isotope exponent of 0.5. Also, models based on a
purely electronic approach cannot capture these effects.

In summary, from measurements of the in-plane magnetic field
penetration depth the zero-temperature superconducting energy gaps
$\Delta_0$ for Y$_{1-x}$Pr$_x$Ba$_2$Cu$_3$O$_{7-\delta}$ ($0.0\leq
x\leq0.45$) were determined. A nontrivial linear relation between
$\Delta_0$ and $T_c$ was found as predicted theoretically. The OIE
on the superconducting gap $\Delta_0$ scales linearly with the one
on $T_c$ and reverses sign around optimum doping. This sign reversal
is in agreement with predictions from model calculations and
unexpected from other theories. Different doping levels of the
isotope exchanged samples were ruled out by performing careful
back-exchange experiments. The experimental results together with
their theoretical analysis suggest that unconventional
electron-lattice interactions play an important role in the physics
of high-temperature superconductivity.

The authors are grateful to K.~Alex~M\"uller for many stimulating
discussions. This work was supported by the Swiss National Science
Foundation, by the K.~Alex M\"uller Foundation and in part by the
EU Project CoMePhS and the NCCR program {\it Materials with Novel
Electronic Properties} (MaNEP) sponsored by the Swiss National
Science Foundation.

\end{document}